\documentclass[aps,prl,twocolumn,groupedaddress]{revtex4-1}

\usepackage{graphicx}
\usepackage{amsfonts}
\usepackage{amsmath}
\usepackage{amssymb}
\usepackage{natbib}
\usepackage{graphicx}
\usepackage{epsf}
\usepackage{epsfig}
\newcommand{\Mg}{$^{24}\text{Mg}^{+}$}
\newcommand{\be}{\begin{equation}}
\newcommand{\ee}{\end{equation}}
\newcommand{\bem}{\begin{multline}}

\begin{document}

%Title of paper
\title{Trapping of Topological-Structural Defects in Coulomb Crystals}

\author{M. Mielenz}
\author{J. Brox}
\author{S. Kahra}
\author{G. Leschhorn}
\author{M. Albert}
\author{T. Schaetz}
\email{tobias.schaetz@physik.uni-freiburg.de}
\affiliation{Albert-Ludwigs-Universit\"at Freiburg, Physikalisches Institut, Hermann-Herder-Strasse 3, 79104 Freiburg, Germany}
\affiliation{Max-Planck-Institut f\"ur Quantenoptik, Hans-Kopfermann-Strasse 1, D-85748 Garching, Germany}
\author{H. Landa}
\author{B. Reznik}
\affiliation{School of Physics and Astronomy, Raymond and Beverly Sackler Faculty of Exact Sciences, Tel-Aviv University, Tel-Aviv 69978, Israel}

\begin{abstract}
We study experimentally and theoretically structural defects which are formed during the transition from a laser cooled cloud to a Coulomb crystal, consisting of tens of ions in a linear radio frequency trap. We demonstrate the creation of predicted topological defects (``kinks") in purely two-dimensional crystals and also find kinks which show novel dynamical features in a regime of parameters not considered before. The kinks are always observed at the center of the trap, showing a large nonlinear localized excitation, and the probability of their occurrence saturates at $\sim 0.5$. Simulations reveal a strong anharmonicity of the kink's internal mode of vibration, due to the kink's extension into three dimensions. As a consequence, the periodic Peierls-Nabarro potential experienced by a discrete kink becomes a globally confining potential, capable of trapping one cooled defect at the center of the crystal.
\end{abstract}


\maketitle
Stable, collective configurations that are nonperturbative, such as  solitons, kinks, vortices, monopoles, and other structural defects, have been extensively studied theoretically and experimentally in a variety of fields in physics, ranging from fluid mechanics, condensed matter, atomic physics, and optics, to high energy physics and cosmology \cite{TopologicalDefects, Rajaraman, dauxois2006physics}. 
These nonlinear solutions are localized, particlelike objects that can propagate without dispersing, and acquire stability which can often be explained by their underlying topological nature. 
Such topological configurations can be formed during a first- or second-order phase transition \cite{Kibble1976, Kibble1980,Zurek1985}. 
Over the last decades, there has been significant interest in the study of topological defects in discrete systems, which can show similar nonlinear phenomena, such as in the Frenkel-Kontorova (FK) model \cite{Braun2004}. 
Discretized solitons are often referred to as kinks, with translation invariance replaced by lattice-periodicity and the kinks, propagating along the lattice, are subject to the so called Peierls-Nabarro (PN) potential \cite{Braun2004}. 
They were experimentally studied, e.g., in waveguide arrays \cite{lederer2001discrete} and proposed with, e.\,g., cold atoms in optical lattices \cite{TrombettoniSmerzi}. 
Discrete defects can be favorable for precise control of their gap-separated, localized modes of oscillation and, in particular, for studying quantum coherence effects \cite{Breathers_in_Josephson_Ladders,wallraff2003quantum} and quantum information prospects \cite{JosephsonBreatherQubits,Marcovitch, Johanning2009}.
\begin{figure}[tb]
	\centering
		\includegraphics[width=3.2in]{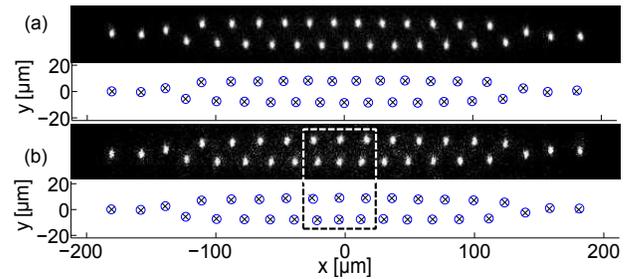}
	\caption{Coulomb crystals, here with 31 ions ($\omega_{\text{z}}/\omega_{\text{y}} \approx 1.34$).
	(a) Top: a CCD image of flourescence light of individual ions trapped and laser cooled in a zigzag configuration. Bottom: circles indicate the fitted positions, while crosses depict the expected positions at the given trapping parameters, derived by MD simulations.
(b) As in (a), however, the ions in the center region forming an extended `kink' defect (dashed box). 
While the ions to the left of the center form a zigzag structure as in (a), the positions on the right are mirrored about the x-axis. 
	}
	\label{fig:figure1}
\end{figure}

In particular, trapped ions \cite{Schneider2012} offer a promising platform for studying the formation and structure of topological defects, as well as for the exploration of their quantum properties. 
When applying laser cooling to trapped ions, they undergo a transition from a state of a chaotic cloud to an ordered structure, a Coulomb crystal \cite{Blumel1988}. 
Isolated from environmental disturbances in an ultrahigh vacuum, 1D chains of ions permit investigating quantum effects at the forefront of quantum metrology \cite{Chou2010}, quantum computing \cite{Schindler2011, Home2009} and quantum simulations \cite{Cirac2012, Porras2004}. 
By adiabatically altering the trapping parameters it is possible to generate phase transitions from one- (linear) to two- (zigzag) and three-dimensional crystals \cite{PhysRevLett.70.818,Piacente2004,MorigiFishman2008structural}. 
Such structural transitions have been proposed to feature quantum phase transitions \cite{Quantum_Zigzag_Transition,BaltruschQuench}.

The experimental creation of localized defects in ion crystals has been achieved during a crystallization of the ion cloud \cite{Schneider2012}. 
It has also been suggested that their creation could be triggered by a nonadiabatic change of the trapping potential \cite{Schneider2012,Landa2010,Campo2010}. 
Realizations of the FK model with ion strings have been suggested in \cite{FKIons,FKIonsHaeffner,FKMorigi}. 
The quantum mechanical properties of kinks in ion traps have been theoretically explored in \cite{Landa2010}, and the connection with the inhomogenous Kibble-Zurek mechanism has been explored in \cite{Campo2010,Chiara2010}. 
Currently, the latter is under investigation, with the creation of kinks reported by quenching the radial \cite{Pyka2012} and the axial \cite{Schneider2012,Ulm2013} confinement.

In this Letter, we present the experimental realization of predicted, extended 2D defects as well as a new class of quasi-3D defects, both formed spontaneously and remaining stable when laser cooling and crystalizing a cloud of ions \footnote{The initially hot ion cloud, including the cooling laser and the ion trap, is an open system and the crystallization is a complex non-adiabatic process, interesting and currently not sufficiently understood by itself.}. 
In the thermodynamic limit, as well as in circular configurations \cite{Landa2010, Schatz2001}, similar discrete solitons (kinks) become topologically protected. 
For the 3D kink we directly observe a highly nonlinear oscillation at its localized, gapped mode, and reveal that the probability to observe a kink at equilibrium shows a saturation at $\sim 0.5$, as the number of ions in the crystal is increased.
We study numerically the structural and dynamical properties of these kinks and find that their spatial extension causes a modification of the Peierls-Nabarro potential, which results in an inherent trapping of the kink at the center of the Coulomb crystal.

The experimental setup consists of a linear Paul trap ($\Omega_{\text{rf}} = 2\pi \cdot 6.22$\,MHz) \cite{Kahra2012, Leschhorn2012} trapping Coulomb crystals of different size of choice (here 10 - 65 photoionized \Mg  ions). 
Experiments were carried out at a single ion trapping frequency of $\omega_{\text{x}} \approx 2\pi \cdot 56$\,kHz (axial direction), while the radial frequencies were varied in the range of $\omega_{\text{y}} \approx 2\pi \cdot 320$\,kHz to $2\pi \cdot 630$\,kHz, corresponding to a total depth of the rf-trapping potential energy of k$_\text{B}$\,10$^4$\,K.
The radial indegeneracy $\omega_{\text{z}}/\omega_{\text{y}}$ is tuned between 1 and 1.4. 
The ions are Doppler cooled via one laser beam, tilted ($\sim$ 5$^{\circ}$) with respect to x-axis. It drives the S$_{1/2}$-P$_{3/2}$ transition ($\lambda$=~280\,nm, natural linewidth $\Gamma \approx 2\pi \cdot $42\,MHz) and is detuned by $\Delta \approx 2 \cdot \Gamma/2$ at a saturation of about 0.2 (Doppler cooling limit T$_{\text{D}}\approx$~1\,mK at $\Delta = \Gamma/2$). 
Data are acquired with a CCD camera detecting fluorescence light of \Mg to determine the ion positions (see figure~\ref{fig:figure1}). 
The resolution is given by the magnified (10 x) pixel size (0.8\,$\mu$m) of the image for 200~ms. 
By decreasing (increasing) the radial (axial) confinement, we reproduce increasingly complex structures.
At constant confinement, the structure and its dimensionality depend on the number of trapped ions, due to the impact of their charge on the total potential.

To calibrate our system and to gain a deeper understanding of the structure and the dynamics within the crystals, we numerically simulate our experimental results by a molecular dynamics (MD) code \cite{KinkMeasurementLong}. 
We take into account the full time-dependent trapping potential and run an optimization routine which considers all trapping parameters and the projection of the crystal plane (xy) on the CCD chip chosen roughly perpendicular to the z-axis. 
We reproduce all observed crystals consistently, with a maximal average deviation of $0.5\mu\text{m}$ per ion.
 We obtain identical results in a simulation using a constant harmonic potential, verifying that micromotion \cite{Landa2012} does not alter our findings.

\begin{figure}[tb]
	\centering
		\includegraphics[width=3.2in]{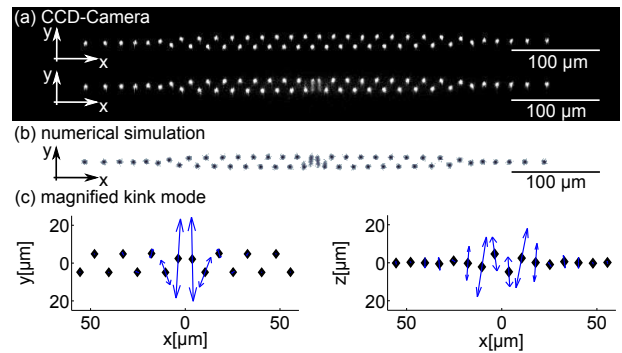}
	\caption{Coulomb crystals containing 50 ions, similar to those depicted in figure~\ref{fig:figure1}, however for reduced asymmetry of the radial confinement, $\omega_{\text{z}}/\omega_{\text{y}}\approx$~1.05. 
(a) CCD images of two crystals obtained for identical experimental parameters. 
Top: Ture zigzag featuring the minimal energy configuration. 
Bottom: Two central ions clearly deviate and show a blurred extension in the radial direction. 
As in figure \ref{fig:figure1}b, the right parts of the crystals are close to identical while the left part of the lower crystal depicts a flipped ``zagzig" structure, separated by the two central ions.
(b) Integrated numerical results for the structure depicted in (a)-Bottom, projected onto the crystal plane. 
(c) Numerically derived components of ion oscillations in the localized, low-frequency normal mode of the kink in (perpendicular to) the crystal plane are depicted on the left (right).}
	\label{fig:KinkBlurring}
\end{figure}

\begin{figure}[tb]
	\centering
		\includegraphics[width=3.2in]{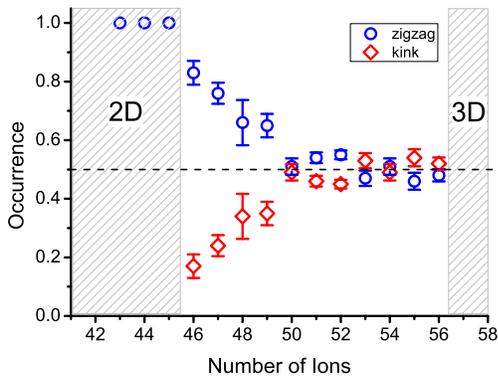}
	\caption{Experimentally derived probability for the occurrence of a pure zigzag and a zigzag with a single kink as depicted in figure~\ref{fig:KinkBlurring}, formed during a nonadiabatic cooling-induced transition from a cloud to a crystal, plotted versus the number of ions. 
With 27 ions and more the minimum energy configuration is a zigzag structure (blue circles). 
Starting with 46 ions, we observe an increasing probability for a zigzag structure with one kink (red rhombi) at the center of the crystal. 
No simultaneous occurrence of two or more defects is observed. 
The probability for observing one kink saturates at $\sim 0.5$ (dashed line) for up to 56 ions. 
The error bars represent the statistical error (1\,$\sigma$) based on the number of attempts.}
	\label{fig:prop_kink}
\end{figure}

In the investigated regime of trapping parameters, for a number of ions between 20 and 60, zigzag structures with and without local defects occur (see figure~\ref{fig:figure1}). 
The duration after which a zigzag structure or a structural defect gets destroyed (e.g. by collision with background gas) amounts to 10 seconds  and exceeds the natural time scale in Coulomb crystals ($\sim 1 / \omega_{\text{x}}$) by five orders of magnitude. 
The two imaged structures in figure~\ref{fig:figure1}, are obtained for $\omega_{\text{z}}/\omega_{\text{y}}\approx$~1.34, where $\omega_{\text{z}}$ is the in-plane and $\omega_{\text{y}}$ is the out-of-plane radial trapping frequency. 
Figure~\ref{fig:figure1}a presents the established zigzag configuration, which is the global minimum of potential energy, while figure~\ref{fig:figure1}b depicts a local minimum, incorporating an extended kink.
The zigzag structures considered so far are extended in two dimensions (xy) only, independent of the presence of a kink. Keeping the axial confinement constant and lowering $\omega_{\text{z}}/\omega_{\text{y}}$, we create qualitatively new defects (figure~\ref{fig:KinkBlurring}a). 
For the rest of this Letter, we discuss their occurrence and properties ($\omega_{\text{z}}/\omega_{\text{y}}\approx$~1.05). 
In this regime, zigzag structures without a kink occur starting with 27 ions, and remain two dimensional with up to 52 ions. 
With 53 ions, a structural phase transition into 3D can be deduced from the numerical results, the space charge of the ions defocusing a few central ions slightly out of the zigzag plane (by $\lesssim$1\,$\mu$m). 
Further increasing the number of ions causes a growing fraction of the ions to extend into the z direction, however, the global minimum configuration (which remains close to a planar zigzag) is still unambiguously identifiable. 
For more than 56 ions, we observe projections of complex structures.

Starting with 46 ions, we detect the random occurrence of local defects mainly involving two ions. The directly observable features of the novel class of kinks are the following. 
(i) They sustain a large motional amplitude of the two center ions along the y-axis, comparable to the radial extension of the crystal, causing the blurring of their images (see figure 2a). 
(ii) The probability for the occurrence of a kink increases with increasing the number of ions, and saturates at $\sim 0.5$ for more than 50 ions (see figure~\ref{fig:prop_kink}). 
(iii) Kinks occur at the center of the crystal only. 
In order to elucidate feature (i), we use MD simulations as described above and reproduce the positions of the ions and their dynamics in the xy plane (see figure~\ref{fig:KinkBlurring}b). 
The crystals including one defect remain nearly identical to a pure zigzag, except at the center. 
There the central ions extend into the transverse (z) direction (see figure~\ref{fig:KinkBlurring}c), even for parameters such that the zigzag crystal without a defect is still planar. 
We further investigate the properties of the kinks numerically applying two separate approaches: (1) a linearization of the oscillations of the ions at small amplitudes and (2) a simulation of the adiabatic dynamics of the kinks as they move along the lattice.

The dispersion relations for the pure zigzag and the crystal with kink (figure~\ref{fig:KinkBlurring}) exhibit for the largest fraction indistinguishable mode spectra. 
However, whereas the lowest mode of the zigzag is the axial center of mass mode (with $\omega_{\text{x}}$), the kink features a gapped mode below $\omega_{\text{x}}$, localized on the center ions. 
A dynamical simulation assuming a thermal energy of k$_\text{B}\text{T}_\text{D}$ for all modes and integrating a ``fluorescence" image from the numerically obtained ion trajectories, reproduces accurately the blurring of the central ions (see figure \ref{fig:KinkBlurring}b). 
We numerically exclude micromotion or the near-axial incidence of the laser at realistic temperatures of $\sim$~5\,T$_\text{D}$ to feature the blurring. 
In fact, it is reproduced exactly, even assuming an excitation of the low-frequency mode only. 
A careful analysis of the dynamics and mode structure \cite{KinkMeasurementLong} reveals that the large amplitude is due to the highly anharmonic oscillation of the ions on the localized low-frequency mode. 
In addition, the frequency of the low energy mode $\omega_\text{low}$ can be controlled by tuning the radial trapping frequencies $\omega_{\text{z}}/\omega_{\text{y}}$ over a range of 0 to  $\sim$~2\,$\omega_{\text{x}}$.

\begin{figure}[tb]
\center {\includegraphics[width=3.in]{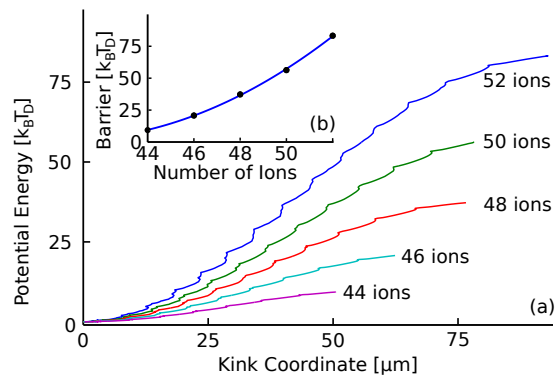}
\caption{Simulation of the effective PN potential energy of a kink (see figure~\ref{fig:KinkBlurring}), (a) dependent on its distance from the center of the Coulomb crystal. 
The potential is symmetric about the trap center and reaches its maximum roughly at the final values shown (dependent on the total ion number). 
(b) The total potential barrier for the kink to escape its trap by reaching the sides is given in dependence of the number of ions, fitted by a quadratic function (blue line). \label{fig:KinkPotential}}
}
\end{figure}

To further elucidate features (ii) and (iii) of the observed kinks, we focus on this mode, that describes the ion oscillations near a minimum of the Peierls-Nabarro (PN) potential. 
For constant ion density (obtainable, e.\,g., in a circular trap \cite{Landa2010, Schatz2001} or in an anharmonic trap \cite{AnharmonicDuan}), the PN potential is periodic with the inter-ion distance. 
In a harmonic trap, the space charge density and the radial displacement decrease out of the center of the crystal. 
If  $\omega_{\text{z}}/\omega_{\text{y}} \gg 1$, the depth of the local minima of the PN potential decreases progressively toward the sides \cite{Campo2010, Chiara2010}, where the kink dissolves into the linear chain, after a random walk \footnote{In the FK model, the shape of the {PN} potential can be obtained numerically, following adiabatically the trajectory of the kink along the lattice \cite{Braun2004}. However, deriving the PN potential becomes highly nontrivial for discrete kinks, as we describe elsewhere \cite{KinkMeasurementLong}.}.

To analyze the motion of the 3D defect along the lattice for our case of $\omega_{\text{z}}/\omega_{\text{y}} \gtrsim 1$, we run a dynamical MD simulation for a kink created initially at an off-centered position. 
We follow its adiabatic dynamics while cooling the motional degrees of freedom of the ions to overdamp the dynamics. 
The kink ``slides" down toward the center of the crystal within an effective trapping potential. 
Tracing the ``downhill" motion of the kink allows to calculate the potential along the adiabatic kink trajectory. 
We reveal that the effective PN potential is no longer periodic, but rather modified to an overall trapping potential (depicted in figure \ref{fig:KinkPotential}a) and a global depth depending quadratically on the total number of ions (see figure \ref{fig:KinkPotential}b). 

For 44 ions, we derive from the simulation the global potential depth for a kink within the crystal to be 10\,k$_\text{B}\text{T}_\text{D}$ (figure \ref{fig:KinkPotential}a). 
This is the height of the barrier for the kink's escape to the sides.
The minimal equilibrium temperature of the ions is a few times T$_{\rm D}$. 
Therefore, during crystallization and formation of the defects, a kink can be created with sufficient energy to escape its trapping potential. 
However, the potential depth for a kink rises with the number of ions such that for 46 ions, it already amounts to $\sim $20\,k$_\text{B}\text{T}_{\rm D}$, permitting us to trap the defect with higher probability. 

Since the minimum configuration has a $\mathbb{Z}_2$ symmetry (of a zigzag and its mirror image), a defect spatially following a kink must be its antikink, and all trapped defects will slide to the center and pairwise annihilate (at given experimental temperature), either leaving one centered kink (if the initial number of kinks was odd) or none. 
Thus, if the probabilities to create an even and an odd number of kinks were exactly equal and independent of the amount of ions, the probability for the final occurrence of one kink would have to be 0.5. 
The fact that we observe a small oscillation of the occurrence around 0.5 (see figure~\ref{fig:prop_kink}), dependent on the number of ions, might be a remainder of the initial statistics for the number of created kinks. A concise analysis will be published elsewhere \cite{KinkMeasurementLong}.
Currently, we are not capable of observing these processes directly, and our simulations reveal a time scale for the kink motion in the crystal shorter by one to two orders of magnitude compared to the current integration time. Observing this motion and the involved dynamics is a future direction for research.
Additionally, it will be interesting to analyze the interplay of thermal effects with the increase in the number of kinks being formed during the crystallization as the number of ions is increased, and to investigate the underlying dynamics.

Entering the quantum regime for large Coulomb crystals and 2D structures by exploiting the kinks' localized properties is another step. 
Cooling the localized gapped mode to the motional ground state and studying its coupling to the bath (of the rest of the modes) would permit to explore mesoscopic open quantum systems.
The controllable trapping parameters offer the possibility to shape the characteristics or mutate different species of the kinks, e.\,g., by tuning the frequency and gap of the localized mode as described above.
The ``kink trap" storing kinks at the center of the Coulomb crystal will be exploited for studying the interaction of different classes of defects (such as a kink combined with a mass defect or a second kink protected by a mass defect, that we have observed \cite{KinkMeasurementLong}). 
These can stabilize side by side without annihilating and are forced into a short range interaction.

\begin{acknowledgments}
TS was funded by the Deutsche Forschungsgemeinschaft (SCHA973), MM, TS and BR acknowledge the support of the European Commission (STREP PICC). BR acknowledges the support of the Israel Science Foundation and the German-Israeli Foundation. 
\end{acknowledgments}

\end{document}